\documentclass[sigconf]{acmart}

\usepackage{booktabs} % For formal tables

\usepackage{graphicx}
\graphicspath{ {figures/} }

\usepackage{caption}
\usepackage{subcaption}
\usepackage{epstopdf}

\usepackage{multirow}
\usepackage{threeparttable}
\usepackage{pbox}
\usepackage{adjustbox}
\usepackage{booktabs}
\usepackage{array}
\usepackage{tabularx}

\usepackage{amsmath}
\usepackage{hyperref}
\usepackage{amsfonts}
\usepackage{xspace}
\usepackage[ruled,vlined]{algorithm2e}
\usepackage{natbib}

\newtheorem{problem}[theorem]{Problem}
\newtheorem{remark}{Remark}

\newcommand{\eg}{\emph{e.g.,}\xspace}
\newcommand{\ie}{\emph{i.e.,}\xspace}

\newcommand{\etal}{\emph{et~al.}\xspace}
\newcommand{\reals}{\ensuremath{\mathbb{R}}\xspace}
\renewcommand{\v}[1]{\ensuremath{\boldsymbol{#1}}\xspace}
\newcommand{\m}[1]{\v{\mathrm{#1}}}
\newcommand{\term}[1]{\emph{#1}}
\renewcommand{\d}{\ensuremath{d}\xspace}
\newcommand{\n}{\ensuremath{n}\xspace}
\newcolumntype{C}{>{\centering\arraybackslash}X}

\newcommand{\myparagraph}[1]{\vspace{0.5em}\noindent\textbf{#1}}

\setcopyright{none}
\acmDOI{}

\acmISBN{}

\acmConference[ ]{ }{ }{ }
\acmYear{}
\copyrightyear{}

\acmPrice{}

\begin{document}
\title{Adequacy of the Gradient-Descent Method for \\
Classifier Evasion Attacks}
%\titlenote{}
%\subtitle{}
%\subtitlenote{}

\author{Yi Han}
%\authornote{}
%\orcid{}
\affiliation{%
  \institution{School of Computing and Information Systems\\
University of Melbourne}
  %\streetaddress{}
  %\city{} 
  %\state{} 
  %\postcode{}
}
\email{yi.han@unimelb.edu.au}

\author{Ben Rubinstein}
%\authornote{}
%\orcid{}
\affiliation{%
  \institution{School of Computing and Information Systems\\
University of Melbourne}
  %\streetaddress{}
  %\city{} 
  %\state{} 
  %\postcode{}
}
\email{benjamin.rubinstein@unimelb.edu.au}

\renewcommand{\shortauthors}{}

\begin{abstract}
Despite the wide use of machine learning in adversarial settings including
computer security, recent studies have demonstrated vulnerabilities to evasion
attacks---carefully crafted adversarial samples that closely resemble 
legitimate instances, but cause misclassification.  In this paper, we examine
the adequacy of the leading approach to generating adversarial samples---the
gradient descent approach. In particular (1) we perform 
extensive experiments on three datasets, MNIST, USPS and Spambase, in order to 
analyse the effectiveness of the gradient-descent method against non-linear support vector
machines, and conclude that carefully reduced kernel smoothness can
significantly increase robustness to the attack; (2) we demonstrate that separated 
inter-class support vectors lead to more secure models, and propose a quantity
similar to margin that can efficiently predict potential susceptibility to
gradient-descent attacks, before the attack is launched; and (3) we design a
new adversarial sample construction algorithm based on optimising the
multiplicative ratio of class decision functions.
\end{abstract}




\maketitle

\section{Introduction}\label{sec:intro}

Recent years have witnessed several demonstrations of machine learning
vulnerabilities in adversarial settings
\cite{dalvi2004adversarial,Lowd2005,Barreno2006,rubinstein2009antidote,bruckner2011stackelberg,Biggio2012,Goodfellow2014,alfeld2016data,li2016data}.
For example, it has been shown that a wide rage of machine learning models---including deep neural networks (DNNs), 
support vector machines (SVMs), logistic regression, 
decision trees and \(k\)-nearest neighbours (\(k\)NNs)---can be easily 
fooled by evasion attacks via adversarial samples \cite{Biggio2013,Goodfellow2014,Moosavi-Dezfooli2016,
Nguyen2015,Papernot2016_transferability,Papernot2016_limitation}.

\begin{figure}
\centering
\hspace*{-1.3cm}
\begin{subfigure}{.3\textwidth}
  \centering
  \includegraphics[width=.4\textwidth]{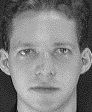}
  \caption{Adam}
  \label{figure_1_a}
\end{subfigure}%
\hspace*{-2.5cm}
\begin{subfigure}{.3\textwidth}
  \centering
  \includegraphics[width=.4\textwidth]{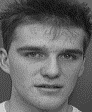}
  \caption{Lucas}
  \label{figure_1_b}
\end{subfigure}%
\hspace*{-2.5cm}
\begin{subfigure}{.3\textwidth}
  \centering
  \includegraphics[width=.4\textwidth]{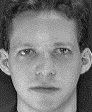}
  \caption{Adam or Lucas?}
  \label{figure_1_c}
\end{subfigure}
\caption{An example evasion attack against a learning model: Image (c)
	is misclassified as Lucas.}
\label{figure_1}
\end{figure}

We refer to the carefully crafted inputs that resemble
legitimate instances but cause misclassification, as \textit{adversarial
samples}, and the malicious behaviours that generate them as
\textit{evasion attacks} \cite{Russu2016}. Figure~\ref{figure_1}
illustrates the attack's effect in the previously-explored vision
domain~\cite{Szegedy2013,Goodfellow2014,Papernot2016_limitation}: %Moosavi-Dezfooli2015
Figures~\ref{figure_1_a} and~\ref{figure_1_b}
present original images from \cite{Samaria1994} for ``Adam'' and ``Lucas'',
who are correctly identified by an SVM face recogniser. However, after human-indiscernible
changes are applied to Figure~\ref{figure_1_a} the model mistakenly identifies
Figure~\ref{figure_1_c} as ``Lucas''.

Thanks to its approximation of best-response
and simplicity, gradient descent has
emerged as the leading approach for creating adversarial samples~\cite{Biggio2013}. 
%\eg in making indiscernible changes
%to an image that leads to misclassification
%\cite{Goodfellow2014}. %,Moosavi-Dezfooli2015,Papernot2016_limitation,Szegedy2013}.
While plentiful in the literature, analyses of evasion attacks tend to present only
a handful of adversarial samples like Figure~\ref{figure_1}, or attack success rates
for limited hyperparameters settings. Such results are useful proofs of concept, but fail
to provide a systematic analysis of the competence of the gradient descent method and best response more generally. Does
the method have any structural limitations? Can gradient descent based 
evasion attacks be reliably thwarted, without cost to learner accuracy? Are there
effective attack alternatives? This paper 
addresses these questions, with a case study on the support vector machine with
radial basis function kernel.

To the best of our knowledge, this is the first rigorous evaluation of 
the gradient descent method in classifier evasion settings. Not only do 
we demonstrate that appropriate choice of smoothness (RBF SVM kernel width) 
can significantly degrade attack success, but that we can even predict 
when attacks will be successful without ever launching them. \\[-0.5em]

\noindent \textbf{Main contributions:} Specifically, our contributions include:
\begin{itemize}
\item An analysis of the kernel precision parameter's impact on the
success rate of evasion attacks, illustrating the existence of 
a phase transition, and concluding that carefully reduced kernel 
smoothness achieves robustness to
gradient-descent attacks without sacrificing SVM accuracy;
\item A novel geometric parameter related to margin,
that strongly correlates with model vulnerability, providing a new avenue
to predict (unseen) attack vulnerability; and
\item A new approach for generating adversarial samples in
multiclass scenarios, with results demonstrating higher 
effectiveness for evasion attacks than the gradient descent method.
\end{itemize}

The remainder of this paper is organised as follows:
Section~\ref{sec:related} overviews previous work on evasion attacks;
Section~\ref{sec:problem} presents our research problem; we present 
a detailed example of how gradient-descent can fail in
Section~\ref{sec:example};  Section~\ref{sec:approach} presents the
gradient-quotient approach for constructing adversarial samples; 
experimental results are presented in Sections~\ref{sec:exp-grad}
and~\ref{sec:exp-quot}; and Section~\ref{sec:conc} concludes the
paper.

\section{Related Work}\label{sec:related}
\citet{Barreno2006} categorise how an
adversary can tamper with a classifier based on whether they have
(partial) control over the training data: in causative attacks, the
adversary can modify the training data to manipulate the learned model; 
in exploratory attacks, the attacker does not poison training, but
carefully alters target test instances to flip classifications. 
See also \cite{Barreno2010,Huang2011}.
This paper focuses on the targeted exploratory case, also known
as evasion attacks~\cite{Biggio2013}. %We summarise previous studies on
%such attacks.

Generalising results on efficient evasion of linear classifiers via reverse
engineering \cite{Lowd2005}, \citet{Nelson2012} consider 
families of convex-inducing classifiers, and propose query algorithms that
require polynomially-many queries and achieve near-optimal modification
cost. The generation of adversarial samples in their setting leverages
membership queries only: the target classifier responds to probes with
signed classifications only. Formulating evasion as optimisation of the 
target classifier's continuous scores, \citet{Biggio2013,Biggio14} first used
gradient descent to produce adversarial samples.

\citet{Szegedy2013} demonstrate changes imperceptible to humans that cause
deep neural networks (DNNs) to misclassify images. Additionally, they offer a 
linear explanation of adversarial samples and design a ``fast gradient sign method'' for
generating such samples \cite{Goodfellow2014}. In a similar vein, 
\citet{Nguyen2015} propose an approach for producing DNN-adversarial 
samples unrecognisable as such to humans.

Papernot \etal published a series of further works in this area:
(1) introducing an algorithm that searches for minimal regions of inputs
to perturb \cite{Papernot2016_limitation}; (2) demonstrating effectiveness
of attacking target models via surrogates---with over 80\% of adversarial
samples launched fooling the victim in one instance \cite{Papernot2016_blackbox};
(3) improved approaches for fitting surrogates, with further investigation
of intra- and cross-technique transferability between DNNs, logistic
regression, SVMs, decision trees and $k$-nearest neighbours 
\cite{Papernot2016_transferability}.

\citet{Moosavi-Dezfooli2015} propose algorithm \textsc{DeepFool} for
generating adversarial samples against DNNs, which leads samples along
trajectories orthogonal to the decision boundary. A similar approach
against linear SVM is proposed in \cite{Papernot2016_transferability}.
Based on \textsc{DeepFool}, \citet{Moosavi-Dezfooli2016} design a method
for computing ``universal perturbations'' that fool multiple DNNs.

In terms of defending against evasion attacks, \citet{Szegedy2013, Goodfellow2014} 
propose injecting adversarial examples into training, in order to improve the 
generalization capabilities of DNNs. This approach resembles active learning. \citet{Papernot2016_distillation} demonstrate 
using distillation strategy against saliency map attack. But it has been proven 
to be ineffective by \cite{Carlini2016_distillation}. 
\citet{Gu2015_architecture, Shaham2016_adversarial, Zhao2016_suppressing, Luo2016_foveation, 
Huang2016_adversary, Carlini2017_robustness, Miyato2016_distributional, Zheng2016_robustness} 
have designed various structural modifications for neural networks.
In addition, \citet{Bhagoji2017_dimensionality, Zhang2016_feature} propose defense methods 
based on dimensionality reduction via principal component analysis (PCA), 
and reduced feature sets, respectively. However, they contradict with \cite{li2014_feature} 
which suggests that more features should be used when facing adversarial evasion.

Most relevant to this paper is the work by \citet{Russu2016}, which
analyses the robustness of SVMs against evasion attacks, including
the selection of the regularisation term, kernel function,
classification costs and kernel parameters. 
%%Ben: removed since it would suggest we should apply our attack against this.
%%It also can suggest--without going into detail--that we've moved on from
%%attacks
%They develop more secure linear and nonlinear classifiers.
Our work delivers a much more detailed analysis of exactly how the
kernel parameters impact vulnerability of RBF SVM, and explanations
of why.

%% Ben:leaving this out, as observation really implied by above, & before
%In summary, as pointed out in \cite{Goodfellow2014}, in many cases modern machine learning techniques ``are not learning the true underlying concepts that determine the correct output label'', which leaves rooms for adversaries to design various ways to manipulate a certain sample, so that it can evade detection, or get mislabelled.

\section{Preliminaries \& Problem Statement}\label{sec:problem}
This section recalls evasion attacks, the gradient-descent method, the RBF SVM,
and summarises the research problem addressed by this paper.

\myparagraph{Evasion Attacks.}
For target classifier $f: \reals^d \to \{-1,1\}$, the purpose of an
\term{evasion attack}
is to apply minimum change \v{\delta} to a target input \v{x}, so that the
perturbed point is misclassified, \ie $\mathrm{sgn}(f(\v{x})) \neq \mathrm{sgn}(f(\v{x} + \v{\delta}))$.
The magnitude of adversarial perturbation \v{\delta} is commonly
quantified in terms of \(L_1\) distance, \ie \(\left\Vert \v{\delta} \right\Vert_1 = \sum_{i=1}^{d} |\delta_{i}|\). 
Formally, evasion attacks are framed as optimisation:
\begin{eqnarray}
	\arg\min_{\v{\delta}\in\reals^d} && \left\Vert \v{\delta} \right\Vert_1 \label{prob-l1}\\
\mbox{s.t.} && \mathrm{sgn}(f(\v{x})) \neq \mathrm{sgn}(f(\v{x}+\v{\delta}))\enspace.\nonumber
\end{eqnarray}

Note that we permit attackers that can modify all features of the input, 
arbitrarily, but that aim to minimise the magnitude of changes. 
Both binary and multiclass scenarios fall into the evasion problem as described;
we consider both learning tasks in this paper.
In multiclass settings, attacks intending to cause specific misclassification
of the test sample are known as \term{mimicry attacks}.

\myparagraph{Gradient-Descent Method.}
The \term{gradient descent method}\footnote{Not to be confused with gradient descent for local optimisation.} has been widely used for generating adversarial
samples for evasion attacks
\cite{Biggio2013,Goodfellow2014,Moosavi-Dezfooli2015,Papernot2016_limitation},
when $f$ outputs confidence scores in \reals and
classifications are obtained by thresholding at $\tau=0$.
The approach applies gradient descent to $f$ directly, initialised at the
target instance. Formally given target instance $\v{x}_0\in\reals^d$ 
evaluating $f(\v{x}_0)>\tau$,
\begin{eqnarray*}
	\v{x}_{t+1} &=& \v{x}_t - \varepsilon_t \cdot \nabla_{\v{x}} f(\v{x}_t)\enspace,
\end{eqnarray*}
where \(\varepsilon_t\) follows an appropriately-selected step size schedule,
and the iteration is terminated when $f(\v{x}_{t})< \tau$.

\myparagraph{The Support Vector Machine.}
Recall the dual program of the soft-margin SVM classifier learner, with hinge-loss

\begin{eqnarray}
	\arg\max_{\v{\alpha}\in\reals^\n} && \v{1}'\v{\alpha} -
\frac{1}{2} \v{\alpha}'\m{G}\v{\alpha}\enspace \\ \label{eq:svm-dual}
\mbox{s.t.} &&
	\v{\alpha}'\v{y}=0,\enspace
	\v{0} \preceq \v{\alpha}\preceq C\v{1} \nonumber
\end{eqnarray}

%\sum_{i,j=1}^{n} \alpha_i\alpha_jy_iy_j

\begin{figure*}[t!]
\begin{minipage}{0.48\textwidth}
  \centering
  \includegraphics[width=1\textwidth]{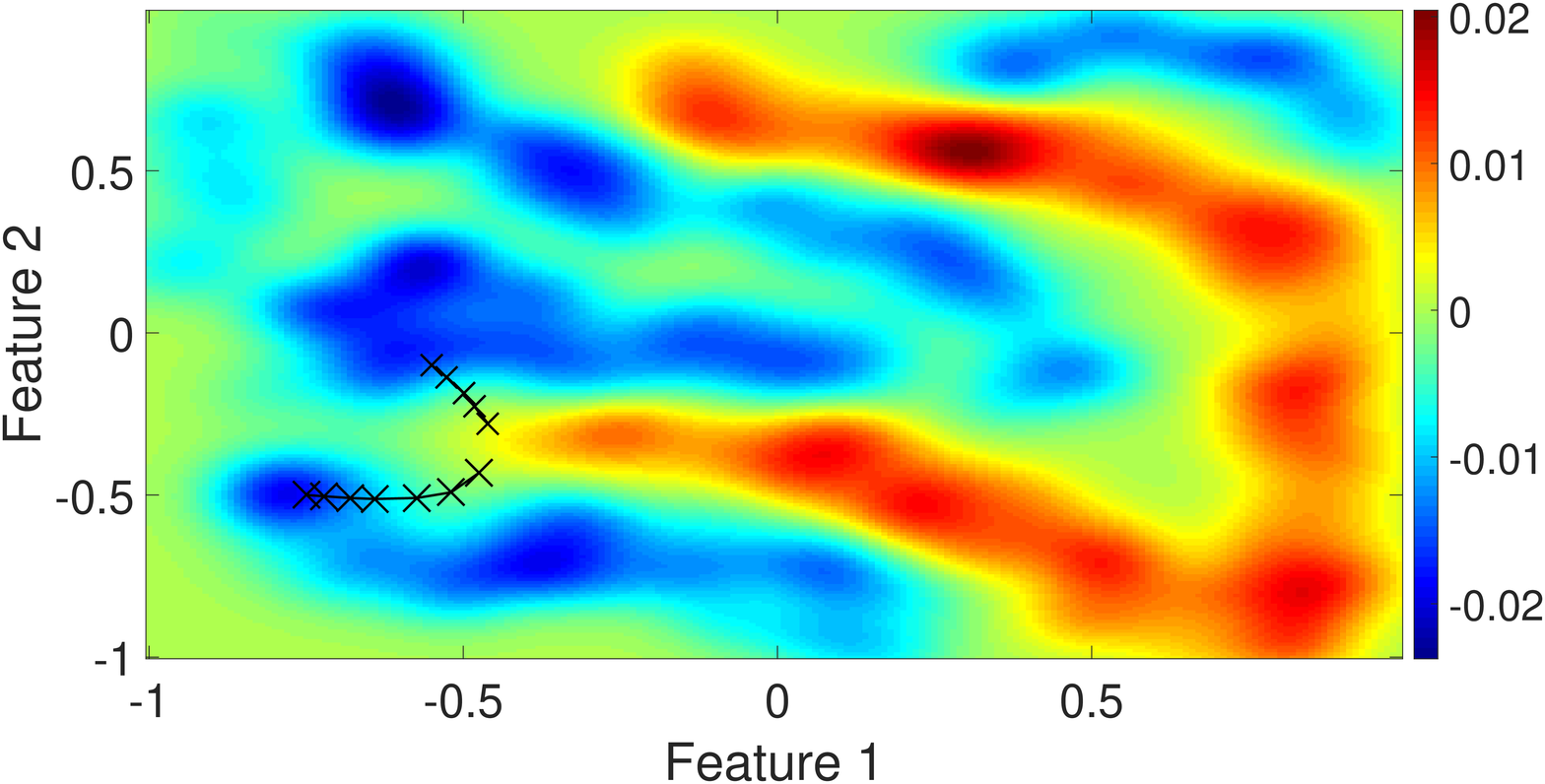} \\[0.5em]
  (a)
  \label{figure_2_a}
\end{minipage}
\hfill
\begin{minipage}{0.5\textwidth}
  \centering
  \includegraphics[width=0.96\textwidth]{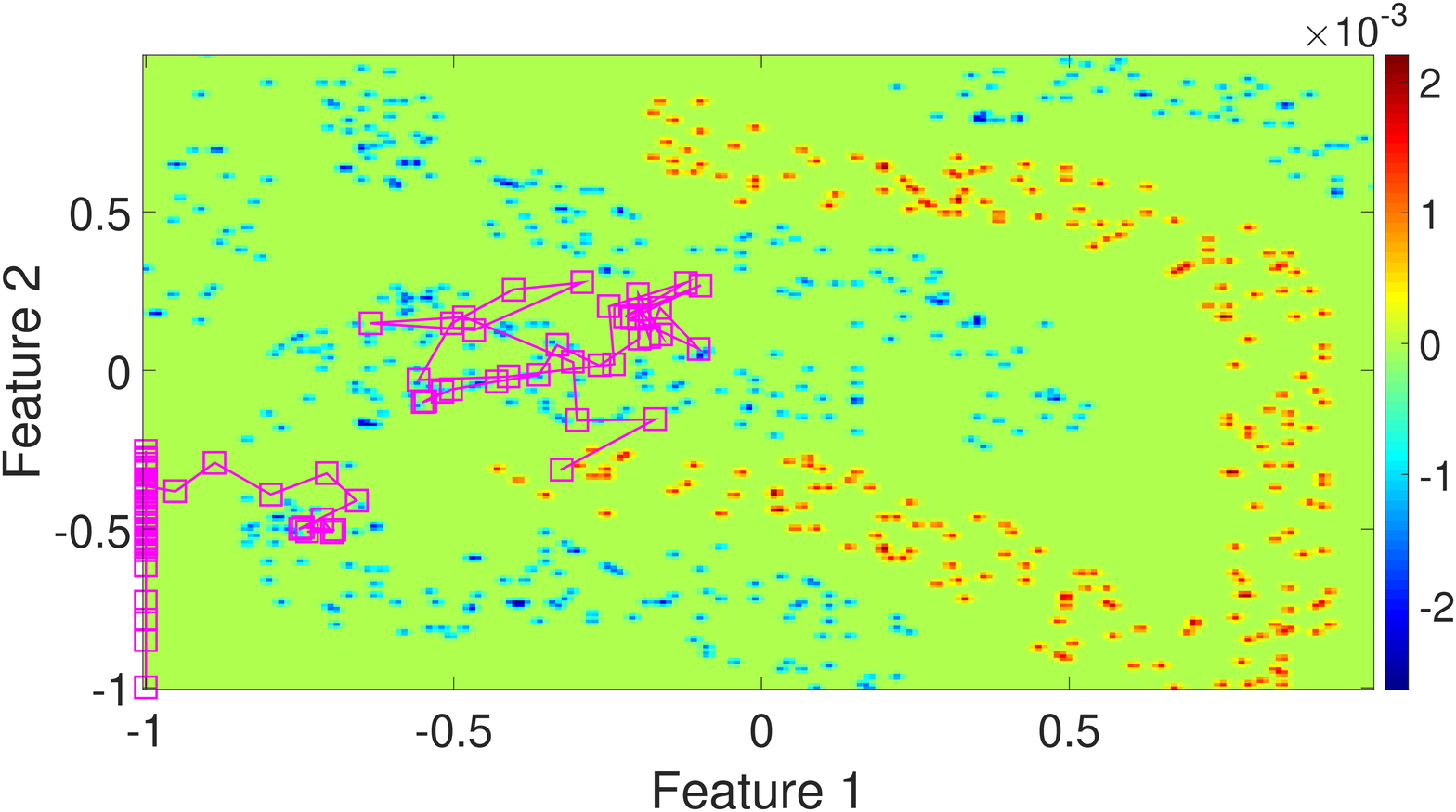} \\[0.5em]
  (b)
  \label{figure_2_b}
\end{minipage}
\caption{Heatmaps visualising decision boundaries of RBF SVM's trained on a sample dataset. (a) \(\gamma=10^2\) model: two black curves display attack paths under gradient descent, of two target points reaching the decision boundary; (b) \(\gamma=10^4\) model: two magenta curves display attack paths under gradient descent, of the two target points now move away from the boundary or take significantly more steps.}
\label{figure_2}
\end{figure*}

where \( \{(\v{x}_i, y_i), i=1, \ldots, \n\} \) is the training data with
\( \v{x}_i \in \reals^{\n} \) and \( y_i \in \{-1, 1\}^{\n} \), \v{\alpha}
are Lagrange multipliers, \v{0}, \v{1} the all zeros, ones vectors, $C>0$
the regularisation penalty parameter on misclassified samples, and
\m{G} the $\n\times \n$ Gram matrix with entries
$G_{ij}= y_i y_j k(\v{x}_i,\v{x}_j)$. The RBF kernel
$k(\v{x}_i,\v{x}_j)=\exp(-\gamma\left\Vert \v{x}_i-\v{x}_j\right\Vert_2^{2})$ has
precision parameter \(\gamma>0\) that
controls kernel width $\sqrt{2/\gamma}$. By the Representer Theorem, 
the learned classifier
\begin{eqnarray}
	f(\v{x}) &=& \sum_{i=1}^{\n}\alpha_i y_i k(\v{x}_i,\v{x})+b\enspace. \label{eq:svm-f}
\end{eqnarray}

\myparagraph{Adequacy and Improvements to Gradient-Descent Method.}
While the gradient-descent method has been effective against a number of machine
learning models, \eg DNNs, linear SVMs, logistic regression
\cite{Biggio2013,Goodfellow2014,Moosavi-Dezfooli2015,Nguyen2015,Papernot2016_blackbox},
there is no general guarantee that gradient descent converges to a global minimum
of $f(\cdot)$ or even converges to local optima quickly---relevant to computational
complexity (a measure of hardness) of evasion. Under linear models---a major focus
of past work---gradient descent quickly finds global optima. While DNNs have been
argued to exhibit local linearity, the existing body of evidence is insufficient
to properly assess the effectiveness of the attack approach. As we argue in the next
section, the approach is in fact \emph{unlikely to be successful against certain
SVMs with RBF kernels, when kernel parameter \(\gamma\) is chosen appropriately}.

\begin{problem}\label{prob:limits}
What limitations of the gradient-descent method are to be expected when 
applied to evasion attacks?
\end{problem}

\begin{problem}\label{prob:defense}
Are there simple defences to the gradient-descent method for popular learners?
\end{problem}

\begin{problem}\label{prob:improvements}
Are there
more effective alternative approaches to generating adversarial samples?
\end{problem}

We address each of these problems in this paper, with special focus on the RBF SVM
as a case study and important example of where the most popular approach to
evasion attack generation can predictably fail, and be improved upon.

\section{Gradient-Descent Method Failure Modes}\label{sec:example}
In this section, we explore how the gradient-descent
method can fail against RBF SVMs with small kernel widths.

\myparagraph{Illustrative Example.}
To demonstrate our key observation, we trained two RBF SVMs on a toy
two-class dataset comprising two features \cite{Chang2011,LIBSVM_url},
using two distinct values for \(\gamma\): \(10^{2}\) and \(10^{4}\).
Figure~\ref{figure_2} displays the heatmaps of the two models' decision 
functions. As can be seen, for the larger \(\gamma\) case, additional
regions result with flat, approximately-zero, decision values. Since the
gradients in these regions are vanishingly small, it is significantly more
likely that an iterate in the gradient-descent method's attack trajectory
will become trapped, or even move towards a direction \emph{away from the
decision boundary altogether}. Notably, both models achieve test accuracies
of 100\%.

In Figure~\ref{figure_2}(a), the two black curves marked with crosses
demonstrate how two initial target points \((-0.55, -0.1)\)
and \((-0.75, -0.5)\), move towards the decision boundary following the
gradient-descent method. However, the two magenta curves marked with squares
in Figure~\ref{figure_2}(b) demonstrate how the same two points either move
away from the boundary or take significantly more steps to reach it, following
the same algorithm but under a different model with a much larger \(\gamma\).

This example illustrates that although the gradient-descent method makes the
test sample less similar to the original class, it does not necessarily
become similar to the other class.

\begin{figure}
\centering
\includegraphics[width=1.0\columnwidth]{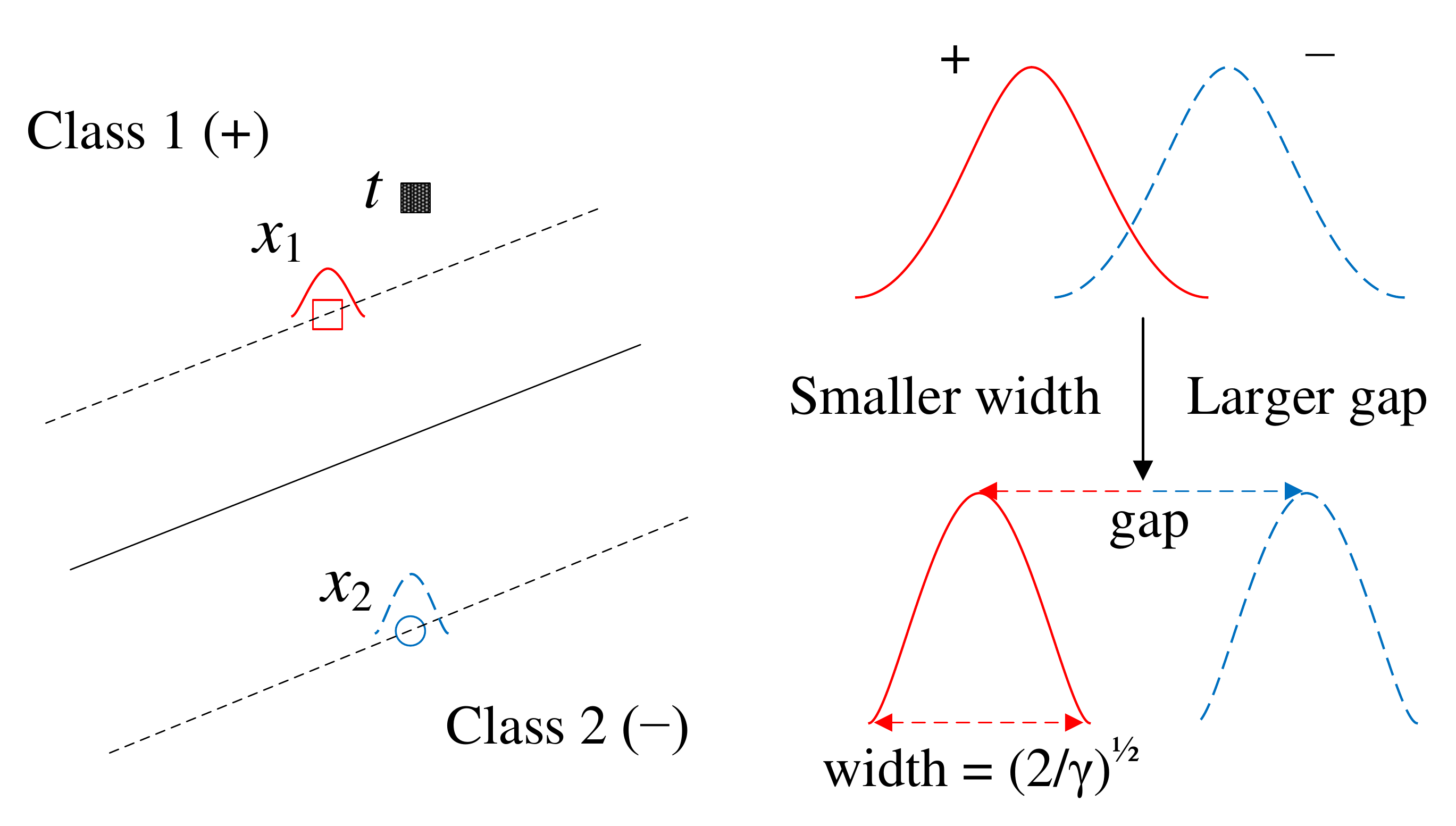}
\caption{Gradient-descent method failure modes.}
\label{figure_3}
\end{figure}

\myparagraph{Discussion.}
We employ Figure~\ref{figure_3} to further explain possible failure modes of
the gradient-descent method: points may get stuck or even move in the wrong
direction. In this given 2D case, \(\v{x}_{1}\) belongs to Class 1, and 
\(\v{x}_{2}\) to Class 2. At instance \v{x}, the \(k^{th} \) component of
the gradient is
\( \nabla^{k} f(\v{x}) = \sum_{i=1}^{2} 2\gamma \alpha_i y_i (x_{i}^{k} - x^{k}) \cdot \exp(-\gamma \left\Vert \v{x} - \v{x}_{i} \right\Vert_2 ^{2}) \).
Clearly, the sign of \( \nabla^{k} \) can be flipped by various choices of
\(\gamma\). Suppose \(x_{1}^{1}, \, x_{2}^{1} > x^{1} \), then solving for
\(\nabla^{1} = 0 \), we obtain the \emph{point at which this phase transition
occurs} as:
\begin{eqnarray*}
	\gamma &=& \frac{\log \alpha_1 (x_{1}^{1} - x^{1}) - \log \alpha_2 (x_{2}^{1} - x^{1})}{\left\Vert \v{x} - \v{x}_{1} \right\Vert_2^{2} - \left\Vert \v{x} - \v{x}_{2} \right\Vert_2^{2}}\enspace.
\end{eqnarray*}

The failure modes hold true in multiclass scenarios. For 
test sample \v{x} the classifier evaluates an $f_i(\v{x})$ per class $i$
and selects the maximiser (a one-vs-all reduction). Suppose that
\(f_{1}(\v{x})\) and \(f_{2}(\v{x})\) are the highest class scores. If
\(\gamma\) is chosen appropriately as above, then gradient descent reduces
both \(f_{1}(\v{x})\) and \(f_{2}(\v{x})\) without ever reranking the two
classes. 

Figure~\ref{figure_3} presents a geometric explanation, where distance
between support vectors of opposite classes exceeding kernel width results
in gradient-descent method iterates becoming trapped in the ``gap'' between.
This section partially addresses Problem~\ref{prob:limits} through the
discussed limitations, while setting $\gamma$ can provide a level of defence
per Problem~\ref{prob:defense}.

\section{The Gradient-Quotient Method}\label{sec:approach}
The previous section motivates Problem~\ref{prob:improvements}'s search for 
effective alternatives to decreasing current class $i$ \(f_{i}(\v{x})\) while
increasing desired class $j$ \(f_{j}(\v{x})\). Rather than moving
in the direction \( -\nabla f_{i}(\v{x}) \) as the gradient-descent method does
(noting this is in the subgradient for the one-vs-all reduction), we propose
following the gradient of the quotient \( - f_{i}(\v{x})/f_{j}(\v{x}) \):
\begin{eqnarray*}
\v{x}_{t+1} &=& \v{x}_t - \varepsilon_t \cdot \nabla (f_{i}(\v{x})/f_{j}(\v{x}))\enspace.
\end{eqnarray*}

\begin{remark}
Employing \( f_{i}(\v{x})-f_{j}(\v{x}) \) in place of
\( f_{i}(\v{x})/f_{j}(\v{x}) \) does not achieve the desired result by the
same flaws suffered by the gradient-descent method:
\(f_{i}(\v{x}), f_{j}(\v{x})\) and \( f_{i}(\v{x})-f_{j}(\v{x}) \)
are decreased simultaneously, while \(f_{i}(\v{x})\) can remain larger than
\(f_{j}(\v{x})\), with no misclassification occurring.
\end{remark}

Note that while in the above, $i$ is taken as the current (maximising) class
index, taking $j$ as the next highest-scoring class corresponds to evasion
attacks while taking $j$ as any fixed target class corresponds to a mimicry 
attack. The results of Section~\ref{sec:exp-quot} establish that this method
can be more effective for manipulating test data in multiclass settings.
However, it is not appropriate to binary-class cases as
\( - f_{1}(\v{x})/f_{2}(\v{x}) = 1\).

\myparagraph{Step Size.}
The step size \(\varepsilon_t\) is important to select carefully: too small
and convergence slows; too large and the attack incurs excessive \(L_{1}\)
change, potentially exposing the attack.

\begin{algorithm}
\LinesNumbered
\SetKwInOut{Input}{Input}\SetKwInOut{Output}{Output}
\Input{Iterate $\v{x}_t$; Current quotient gradient $\nabla\in\reals^{\d}$; Parameter $\eta>0$}
\Output{Step size $\varepsilon_t$}
\BlankLine
Select $i\in\arg\max_{j\in[\d]} |\nabla^j|$.

Select $\varepsilon_t>0$ such that $\varepsilon_t\cdot |\nabla^i|\in\sqrt{t} [5\eta, 10\eta]$.

%\textbf{Return} $\varepsilon$. 
\caption{Gradient-quotient step size.\label{algo:step}}
\end{algorithm}

In our experiments, %instead of using a fixed step size that has to be fine-tuned in each case,
we limit the largest change made to a single feature per iteration, and
determine the step size accordingly as described in Algorithm~\ref{algo:step}.
Here \(\eta\) is a domain-specific value corresponding to a unit change in
a feature, \eg for a grayscale image \(\eta=1\) corresponds to a unit change
intensity level. The select rule's \( [5\eta, 10\eta] \) is motivated by 
round-off practicalities in steps: if the largest gradient component were
smaller than \(5\eta\), it is likely that most other components would be
0, making convergence extremely slow. 
Since \( [5\eta, 10\eta] \) is a relatively conservative start, we increase
it gradually; as explained next, the maximum step number in our experiments
is 30. This increasing step size corresponds to a variant of guess-then-double.
In addition, another advantage of Algorithm~\ref{algo:step} is that it produces a similar 
amount of \(L_{1}\) changes to the test samples, despite different values 
of \(\gamma\).

%is that since \( \nabla_{i} \) needs to be rounded off in each step, 

%(1) at the \(i^{th}\) step, sort all components of the gradient, i.e., \( \nabla_{i} = (\nabla_{i}^{1}, \nabla_{i}^{2}, ..., \nabla_{i}^{j}, ...) \) with \( \nabla_{i}^{j} \) corresponding to the \(j^{th}\) feature, by their absolute values; (2) adjust the largest component \( \nabla_{i}^{k} \), so that \( |\nabla_{i}^{k}| \in \sqrt{i} \cdot [5\eta, 10\eta] \), where \(\eta\) is one unit value, e.g., for a grayscale image, \(\eta\) = 1; (3) update the other components with the same proportion. 

\section{Experiments: Gradient-Descent Method}\label{sec:exp-grad}
Section~\ref{sec:example} demonstrates that RBF SVM is less vulnerable to 
evasion attacks when \(\gamma\) is set appropriately. In this section, we
present a more detailed analysis of the impact of \(\gamma\) on the
attack's success rate, further addressing Problems~\ref{prob:limits}
and~\ref{prob:defense}.

\subsection{Datasets}
Three datasets are selected to facilitate the analysis on \(\gamma\): 
(1) MNIST \cite{LeCun1998,MNIST_url} is a dataset of handwritten digits that has been widely studied. 
It contains a training set of \(6 \times 10^{4} \) samples (only the first
\(5 \times 10^{4} \) are used in our experiments), and a test set of \(10^{4} \) samples. 
Following the approach of \cite{Papernot2016_transferability}, we divide the training set into 
5 subsets of \(10^{4} \) samples, \(D_{1} \sim D_{5}\). Each subset is used to train
models separately. Specifically, in the experiments that study binary-class
settings, only the data in \(D_{1}\) (or the test dataset) that belong to the
two classes are used for training (or testing respectively).
(2) USPS \cite{USPS_url} is another dataset of (normalised) handwritten digits 
scanned from envelopes by the U.S. Postal Service. There are 7291 training observations 
and 2007 test observations, each of which is a 16 \(\times\) 16 grayscale image.
(3) Spambase \cite{Spambase_url} contains a total number of 4601 emails, 
with 1813 (2788) being spam (non-spam). In our experiments, we randomly select 3000 emails
as the training set, and the remainder as the test set. Table~\ref{table_datasets} summarises the 
key information of these three datasets. Note that MNIST and Spambase are scaled to [0,1], 
while USPS is scaled to [-1, 1].

\begin{table}\small
\centering
\caption{Summary of the datasets used in this paper}
\label{table_datasets}
\setlength\extrarowheight{2pt}
\begin{tabular}{ |c|c|c|c|c|c| }
\hline
\multirow{2}{*}{Name} & \multicolumn{2}{c|}{Size} & \multirow{2}{*}{Features} & \multirow{2}{*}{Domain} & \multirow{2}{*}{Range}\\
\cline{2-3}
& Training set & Test set & & &\\
\hline
MNIST & \(5 \times 10^{4} \) & \(10^{4} \) & \(28 \times 28 \) & Vision & [0, 1]\\
\hline
USPS & 7291 & 2007 & \(16 \times 16 \) & Vision & [-1, 1]\\
\hline
Spambase & 3000 & 1601 & 57 & Spam & [0, 1]\\
\hline
\end{tabular}
\end{table}

\begin{figure*}[t!]
\begin{minipage}{0.48\textwidth}
  \centering
  \includegraphics[width=1\columnwidth]{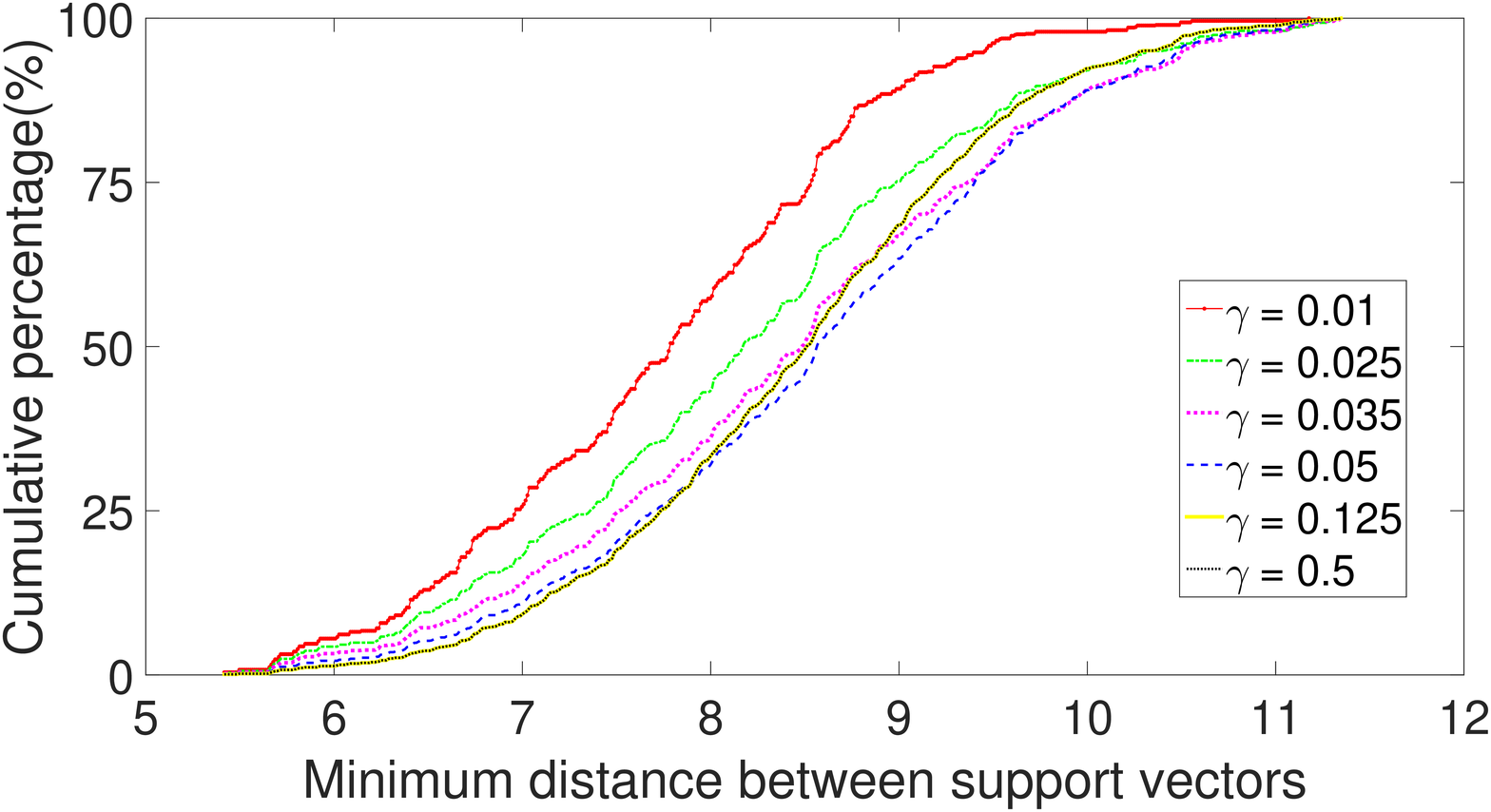}
  \caption{Minimum distance between each support vector of class ``3'' and all support vectors of class ``4'' in the MNIST dataset.}
  \label{figure_4_a}
\end{minipage}\hfill
\begin{minipage}{0.48\textwidth}
  \centering
  \includegraphics[width=1\columnwidth]{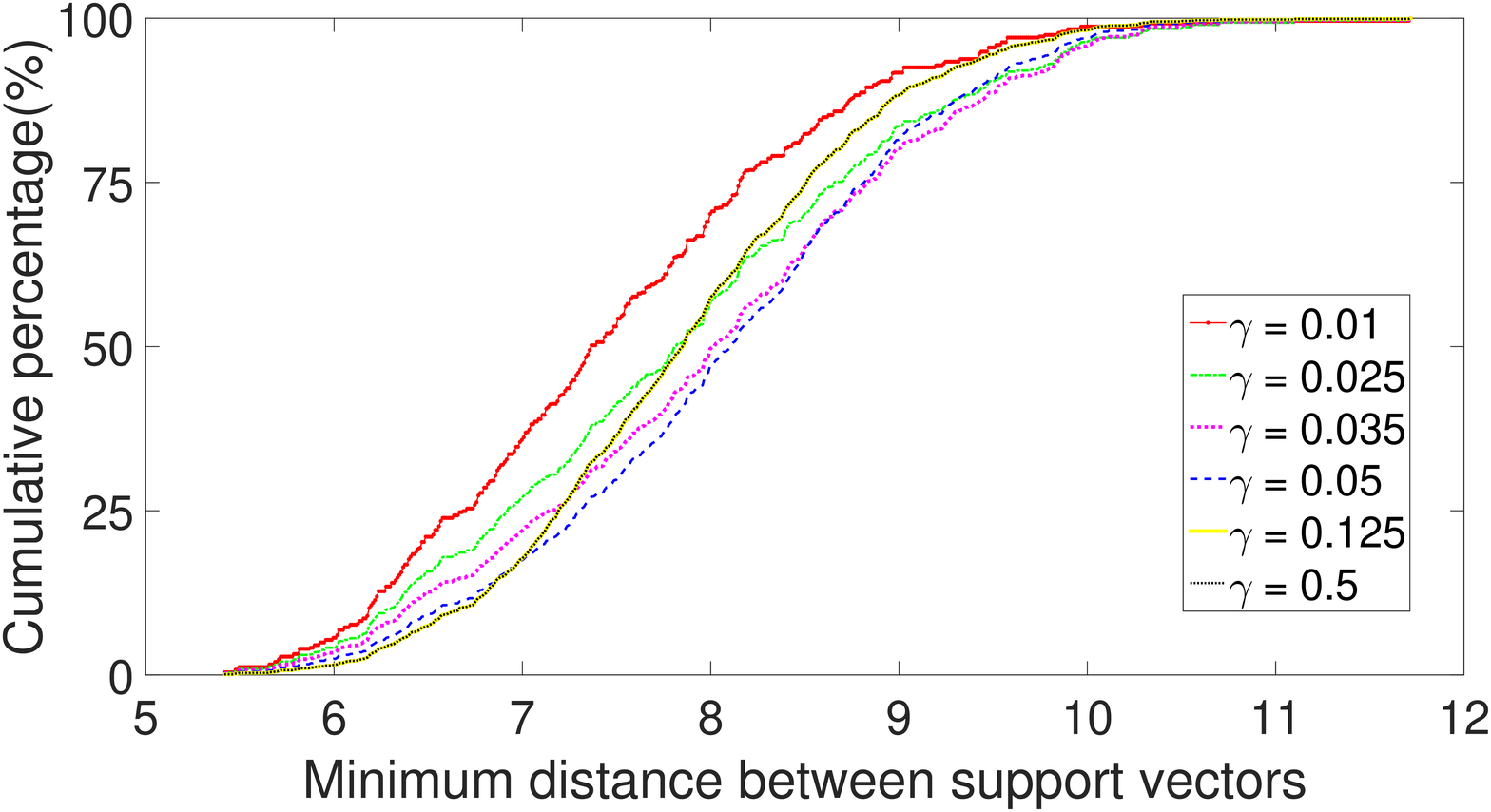}
  \caption{Minimum distance between each support vector of class ``4'' and all support vectors of class ``3'' in the MNIST dataset.}
  \label{figure_4_b}
  \end{minipage}
\end{figure*}

\subsection{Impact of \(\gamma\) on Vulnerability (Binary Class)}
We begin with the binary scenario and investigate how \(\gamma\) impacts the
success rate of causing SVMs to misclassify (1) three pairs of digits---1 \& 2, 3 \&
4, 5 \& 7 for the MNIST dataset, (2) three pairs of digits---0 \& 1, 2 \& 3, 6 \& 7 
for the USPS dataset, and (3) spam and non-spam emails for the Spambase dataset. 
Note that in this subsection, all models regarding the MNIST dataset are trained on
\(D_{1}\). An attack is considered successful if the perturbed test sample is
misclassified within 30 steps. The reason why we choose 30 is that although
larger values will increase the attack's success rate, the changes made to the
original samples are so obvious they would be easily detected by manual audit.

\begin{figure*}
\begin{minipage}{0.33\textwidth}
  \centering
  \includegraphics[width=1\textwidth]{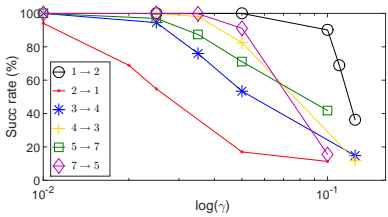} \\
  (a) MNIST
  \label{figure_phase_mnist}
\end{minipage}
\hfill
\begin{minipage}{0.33\textwidth}
  \centering
  \includegraphics[width=1\textwidth]{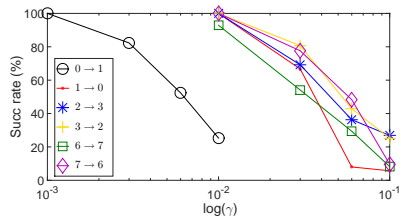} \\
  (b) USPS
  \label{figure_phase_usps}
\end{minipage}
\begin{minipage}{0.33\textwidth}
  \centering
  \includegraphics[width=1\textwidth]{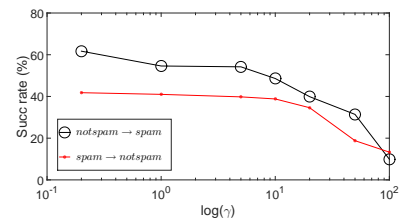} \\
  (c) Spambase
  \label{figure_phase_mnist}
\end{minipage}
\hfill
\caption{Phase transition of the success rate for gradient-descent method evasion attacks.}
\label{figure_phase}
\end{figure*}

Tables~\ref{table_1_a}--\ref{table_1_c},~\ref{table_1_e}--\ref{table_1_g} and~\ref{table_1_spambase} 
illustrate a \textbf{phase transition in all cases: a small decrease of \(\gamma\) causes
a significant jump in success rate}. Figure~\ref{figure_phase} visualises the phase transition. 
\footnote{Testing for effect of
\(C\) revealed much less impact on success rate. Therefore, a smaller value of 
\(C\) is used in certain cases.}

From the attacker's point of view, in addition to minimising the 
overall changes, it is also desirable that no single pixel is modified by a large amount 
when perturbing an image, \ie \(\left\Vert \v{\delta} \right\Vert_{\infty} < B_{\infty} \in R^{+}\) 
(\(B_{\infty}\) can be considered as the attacker's budget). 
Therefore, we modify the original optimisation problem~\ref{prob-l1} as follows:

\begin{eqnarray}
	\arg\min_{\v{\delta}\in\reals^d} && \left\Vert \v{\delta} \right\Vert_1  \label{prob-l1-linf} \\
	\mbox{s.t.} && \mathrm{sgn}(f(\v{x})) \neq \mathrm{sgn}(f(\v{x}+\v{\delta})) \nonumber \\
		    && \left\Vert \v{\delta} \right\Vert_1 < B_{1} \in R^{+},  \nonumber \\
      && \left\Vert \v{\delta} \right\Vert_{\infty} < B_{\infty} \in R^{+} \nonumber
\end{eqnarray}

We apply much stricter stopping criteria, and re-run the experiments on MNIST and USPS. 
%with \(B_{1}\) and \(B_{\infty}\) set to 39.2 (\ie \(10^{4}\) pixels) and 0.3, respectively.
As can be seen from Tables~\ref{table_1_mnist_limit} \& \ref{table_1_usps_limit}, 
a phase transition still exists in each case.

\begin{table*}[t!]\small
\centering
\caption{Success rate and average \(L_{1}\) change for gradient-descent method evasion attacks (binary class, MNIST).}
\label{table_1_mnist}

\begin{subtable}{\textwidth}
\centering
\caption{RBF SVM: MNIST, digits 1 and 2.\label{table_1_a}}
\vspace{-6pt}
\setlength\extrarowheight{2pt}
\begin{tabular}{ |c|c|c|c|c|c|c|c|c|c| }
\hline
\multicolumn{2}{|c|}{\(\gamma\)} & 0.01 & 0.02 & 0.025 & 0.05 & 0.1 & 0.11 & 0.125 & 0.5\\ 
\hline
\multicolumn{2}{|c|}{\(C\)} & \(5 \times 10^{4} \) & \(5 \times 10^{4} \) & \(5 \times 10^{4} \) & \(5 \times 10^{4} \) & \(5 \times 10^{4} \) & \(5 \times 10^{4} \) & \(5 \times 10^{4} \) & 10\\ 
\hline
\multicolumn{2}{|c|}{Accuracy(\%)} & 99.4 & 99.5 & 99.6 & 99.5 & 99.8 & 99.7 & 99.5 & 98.6\\ 
\hline
\multirow{2}{*}{\( 1 \to 2 \)} & Succ rate (\%) & 100 & 100 & 100 & 100 & 90.1 & 68.9 & 36.3 & 1.5\\
\cline{2-10}
&\(P(2/\sqrt{\gamma} \geq Min Dist) \) (\%) & 100 & 100 &100 &100 & 68.4 & 55.0 & 35.0 & 0.1\\
\hline
\multirow{2}{*}{\( 2 \to 1 \)} & Succ rate (\%) & 94.1 & 68.9 & 54.8 & 17.0 & 11.3 & 13.3 & 15.2 & 12.9\\
\cline{2-10}
&\(P(2/\sqrt{\gamma} \geq Min Dist) \) (\%) & 100 & 100 &100 & 59.3 & 7.5 &4.8 & 2.8 & 0.1\\
\hline
\end{tabular}
\end{subtable}

\vspace{14pt}
\begin{subtable}{\textwidth}
\centering
\caption{RBF SVM: MNIST, digits 3 and 4.\label{table_1_b}}
\vspace{-6pt}
\setlength\extrarowheight{2pt}
\begin{tabular}{ |c|c|c|c|c|c|c|c| } 
\hline
\multicolumn{2}{|c|}{\(\gamma\)} & 0.01 & 0.025 & 0.035 & 0.05 & 0.125 & 0.5\\ 
\hline
\multicolumn{2}{|c|}{\(C\)} & \( 10^{4} \) & \( 10^{4} \) & \( 10^{4} \) & \( 10^{4} \) & \( 10^{4} \) & 10\\ 
\hline
\multicolumn{2}{|c|}{Accuracy(\%)} & 99.5 & 99.7 & 99.6 & 99.5 & 99.5 & 99.8\\ 
\hline
\multirow{2}{*}{\( 3 \to 4 \)} & Succ rate (\%) & 100 & 94.5 & 76.0 & 53.4 & 14.8 & 5.3\\
\cline{2-8}
&\(P(2/\sqrt{\gamma} \geq Min Dist) \) (\%) & 100 & 100 & 96.7 & 61.7 & 0.29 & 0.1\\
\hline
\multirow{2}{*}{\( 4 \to 3 \)} & Succ rate (\%) & 100 & 100 & 98.1 & 82.5 & 11.8 & 1.6\\
\cline{2-8}
&\(P(2/\sqrt{\gamma} \geq Min Dist) \) (\%) & 100 & 100 & 99.2 & 79.7 & 0.4 & 0.1\\
\hline
\end{tabular}
\end{subtable}

\vspace{14pt}
\begin{subtable}{\textwidth}
\centering
\caption{RBF SVM: MNIST, digits 5 and 7.\label{table_1_c}}
\vspace{-6pt}
\setlength\extrarowheight{2pt}
\begin{tabular}{ |c|c|c|c|c|c|c|c| }
\hline
\multicolumn{2}{|c|}{\(\gamma\)} & 0.01 & 0.025 & 0.035 & 0.05 & 0.1 & 0.5\\ 
\hline
\multicolumn{2}{|c|}{\(C\)} & \( 10^{4} \) & \( 10^{4} \) & \( 10^{4} \) & \( 10^{4} \) & \( 10^{4} \) & 10\\ 
\hline
\multicolumn{2}{|c|}{Accuracy(\%)} & 99.3 & 99.5 & 99.6 & 99.6 & 99.1 & 99.7\\ 
\hline
\multirow{2}{*}{\( 5 \to 7 \)} & Succ rate (\%) & 100 & 97.0 & 87.4 & 71.1 & 41.8 & 6.2\\
\cline{2-8}
&\(P(2/\sqrt{\gamma} \geq Min Dist) \) (\%) & 100 & 100 & 98.7 & 75.6 & 0.2 & 0.1\\
\hline
\multirow{2}{*}{\( 7 \to 5 \)} & Succ rate (\%) & 100 & 100 & 99.8 & 91.1 & 15.5 & 2.3\\
\cline{2-8}
&\(P(2/\sqrt{\gamma} \geq Min Dist) \) (\%) & 100 & 100 & 100 & 87.2 & 0.2 & 0.1\\
\hline
\end{tabular}
\end{subtable}

\vspace{14pt}
\begin{subtable}{\textwidth}
\centering
\caption{Linear SVM.\label{table_1_d}}
\vspace{-6pt}
\setlength\extrarowheight{2pt}
\begin{tabular}{ |c|c|c|c|c|c|c| } 
\hline
\multirow{2}{*}{} & \multicolumn{3}{c|}{\(C=1000\)} & \multicolumn{3}{c|}{\(C=5000\)}\\
\cline{2-7}
&  Succ rate (\%) & Ave \(L_{1}\) change & Accuracy (\%) & Succ rate (\%) & Ave \(L_{1}\) change & Accuracy (\%)\\
\hline
\( 1 \to 2 \) & 100 & 22.8 & \multirow{2}{*}{99.3} & 100 & 18.4 & \multirow{2}{*}{99.2}\\
\cline{1-3}\cline{5-6}
\( 2 \to 1 \) & 100 & 37.5 & & 100 & 34.8 & \\
\hline
\( 3 \to 4 \) & 100 & 39.5 & \multirow{2}{*}{99.7} & 100 & 37.3 & \multirow{2}{*}{99.6}\\
\cline{1-3}\cline{5-6}
\( 4 \to 3 \) & 100 & 32.8 & & 100 & 25.9 & \\
\hline
\( 5 \to 7 \) & 100 & 30.4 & \multirow{2}{*}{99.0} & 100 & 28.1 & \multirow{2}{*}{99.0}\\
\cline{1-3}\cline{5-6}
\( 7 \to 5 \) & 100 & 29.1 & & 100 & 23.6 & \\
\hline
\end{tabular}
\end{subtable}
\end{table*}

\vspace{14pt}

\begin{table*}[t!]\small
\centering
\caption{Success rate for gradient-descent method evasion attacks (binary class, USPS).}
\label{table_1_usps}

\vspace{-4pt}
\begin{subtable}{\textwidth}
\centering
\caption{RBF SVM: USPS, digits 0 and 1.\label{table_1_e}}
\vspace{-6pt}
\setlength\extrarowheight{2pt}
\begin{tabular}{ |c|c|c|c|c|c|c|c|c| }
\hline
\multicolumn{2}{|c|}{\(\gamma\)} & 0.001 & 0.003 & 0.006 & 0.01 & 0.03 & 0.06 & 0.1\\ 
\hline
\multicolumn{2}{|c|}{\(C\)} & \(10^{4} \) & \(10^{4} \) & \(10^{4} \) & \(5 \times 10^{3} \) & \(5 \times 10^{3} \) & \(5 \times 10^{3} \) & \(10^{3} \)\\ 
\hline
\multicolumn{2}{|c|}{Accuracy(\%)} & 99.2 & 99.5 & 99.4 & 99.5 & 99.5 & 99.4 & 99.4\\ 
\hline
\multirow{2}{*}{\( 0 \to 1 \)} & Succ rate (\%) & 100 & 82.3 & 52.4 & 25.3 & 4.7 & 3.4 & 3.6\\
\cline{2-9}
&\(P(2/\sqrt{\gamma} \geq Min Dist) \) (\%) & 100 & 100 &100 &100 & 35.7 & 6.0 & 0.6\\
\hline
\multirow{2}{*}{\( 1 \to 0 \)} & Succ rate (\%) & 100 & 100 & 100 & 100 & 66.7 & 8.0 & 5.7\\
\cline{2-9}
&\(P(2/\sqrt{\gamma} \geq Min Dist) \) (\%) & 100 & 100 & 100 & 100 & 100 & 98.3 & 22.0\\
\hline
\end{tabular}
\end{subtable}

\vspace{6pt}
\begin{subtable}{\textwidth}
\centering
\caption{RBF SVM: USPS, digits 2 and 3.\label{table_1_f}}
\vspace{-6pt}
\setlength\extrarowheight{2pt}
\begin{tabular}{ |c|c|c|c|c|c| } 
\hline
\multicolumn{2}{|c|}{\(\gamma\)} & 0.01 & 0.03 & 0.06 & 0.1\\ 
\hline
\multicolumn{2}{|c|}{\(C\)} & \( 10^{5} \) & \( 10^{5} \) & \( 10^{5} \) & \( 10^{5} \)\\ 
\hline
\multicolumn{2}{|c|}{Accuracy(\%)} & 97.5 & 97.8 & 97.8 & 97.8\\ 
\hline
\multirow{2}{*}{\( 2 \to 3 \)} & Succ rate (\%) & 100 & 69.2 & 36.3 & 26.9\\
\cline{2-6}
&\(P(2/\sqrt{\gamma} \geq Min Dist) \) (\%) & 100 & 97.9 & 45.4 & 4.6\\
\hline
\multirow{2}{*}{\( 3 \to 2 \)} & Succ rate (\%) & 100 & 80.7 &42.9 & 25.2\\
\cline{2-6}
&\(P(2/\sqrt{\gamma} \geq Min Dist) \) (\%) & 100 & 100 & 52.1 & 11.1\\
\hline
\end{tabular}
\end{subtable}

\vspace{6pt}
\begin{subtable}{\textwidth}
\centering
\caption{RBF SVM: USPS, digits 6 and 7.\label{table_1_g}}
\vspace{-6pt}
\setlength\extrarowheight{2pt}
\begin{tabular}{ |c|c|c|c|c|c| } 
\hline
\multicolumn{2}{|c|}{\(\gamma\)} & 0.01 & 0.02 & 0.03 & 0.06\\ 
\hline
\multicolumn{2}{|c|}{\(C\)} & \( 10^{4} \) & \( 10^{4} \) & \( 10^{4} \) & \( 10^{4} \)\\ 
\hline
\multicolumn{2}{|c|}{Accuracy(\%)} & 99.7 & 99.7 & 99.7 & 99.7\\ 
\hline
\multirow{2}{*}{\( 6 \to 7 \)} & Succ rate (\%) & 92.9 & 54.1 & 29.4 & 8.2\\
\cline{2-6}
&\(P(2/\sqrt{\gamma} \geq Min Dist) \) (\%) & 100 & 99.2 & 65.9 & 19.4\\
\hline
\multirow{2}{*}{\( 7 \to 6 \)} & Succ rate (\%) & 100 & 77.6 & 48.3 & 9.6\\
\cline{2-6}
&\(P(2/\sqrt{\gamma} \geq Min Dist) \) (\%) & 100 & 100 & 86.5 & 24.3\\
\hline
\end{tabular}
\end{subtable}
\end{table*}

\vspace{6pt}
\begin{table*}\small
\centering
\caption{Success rate for gradient-descent method evasion attacks (binary class, Spambase).}
\vspace{-6pt}
\label{table_1_spambase}
\begin{threeparttable}
\setlength\extrarowheight{2pt}
\begin{tabular}{ |c|c|c|c|c|c|c|c|c| }
\hline
\multicolumn{2}{|c|}{\(\gamma\)} & 0.2 & 1 & 5 & 10 & 20 & 50 & 100\\ 
\hline
\multicolumn{2}{|c|}{\(C\)} & \(5 \times 10^{5} \) & \(5 \times 10^{5} \) & \(10^{5} \) & \(10^{5} \) & \(10^{5} \) & \(10^{5} \) & \(5 \times 10^{4} \)\\ 
\hline
\multicolumn{2}{|c|}{Accuracy(\%)} & 91.3 & 92.9 & 92.9 & 92.9 & 92.6 & 92.1 & 91.4\\ 
\hline
\multirow{2}{*}{\( 0 \to 1 \)\textsuperscript{a}} & Succ rate (\%) & 61.7 & 54.6 & 54.2 & 48.6 & 39.9 & 31.4 & 9.9\\
\cline{2-9}
&\(P(2/\sqrt{\gamma} \geq Min Dist) \) (\%) & 100 & 99.9 & 94.5 & 86.8 & 73.3 & 53.6 & 41.8\\
\hline
\multirow{2}{*}{\( 1 \to 0 \)} & Succ rate (\%) & 41.8 & 41.0 & 39.8 & 38.8 & 34.6 & 18.8 & 13.3\\
\cline{2-9}
&\(P(2/\sqrt{\gamma} \geq Min Dist) \) (\%) & 100 & 100 & 97.7 & 94.6 & 89.4 & 64.6 & 41.7\\
\hline
\end{tabular}
\begin{tablenotes}
\footnotesize
\item[a] 0: not spam, 1: spam
\end{tablenotes}
\end{threeparttable}
\end{table*}

\begin{table}[h!]\small
\centering
\caption{Success rate of gradient-descent method evasion attacks
(binary class, MNIST, with \(L_{1}\) and \(L_{\infty}\) limit)}
\label{table_1_mnist_limit}

\vspace{-4pt}
\begin{subtable}{0.5\textwidth}
\centering
\caption{digits 1 and 2, \(B_{1}\)=39.2, \(B_{\infty}\)=0.3\label{table_1_h}}
\vspace{-6pt}
\setlength\extrarowheight{2pt}
\begin{tabular}{ |c|c|c|c|c| }
\hline
\multicolumn{2}{|c|}{\(\gamma\)} & 0.01 & 0.05 & 0.1\\
\hline
\multicolumn{2}{|c|}{\(C\)} & \(5 \times 10^{4} \) & \(5 \times 10^{4} \) & \(5 \times 10^{4} \)\\
\hline
\multicolumn{2}{|c|}{Accuracy (\%)} & 99.4 & 99.5 & 99.8\\
\hline
\multirow{2}{*}{Succ rate (\%)} & \( 1 \to 2 \) & 99.5 & 92.7 & 30.4\\
\cline{2-5}
 & \( 2 \to 1 \) & 27.2 & 8.4 & 9.8\\
\hline
\end{tabular}
\end{subtable}

\vspace{6pt}
\begin{subtable}{0.5\textwidth}
\centering
\caption{digits 3 and 4, \(B_{1}\)=39.2, \(B_{\infty}\)=0.3\label{table_1_i}}
\vspace{-6pt}
\setlength\extrarowheight{2pt}
\begin{tabular}{ |c|c|c|c|c| }
\hline
\multicolumn{2}{|c|}{\(\gamma\)} & 0.01 & 0.05 & 0.125\\
\hline
\multicolumn{2}{|c|}{\(C\)} & \(10^{4} \) & \(10^{4} \) & \(10^{4} \)\\
\hline
\multicolumn{2}{|c|}{Accuracy (\%)} & 99.5 & 99.5 & 99.5\\
\hline
\multirow{2}{*}{Succ rate (\%)} & \( 3 \to 4 \) & 34.5 & 18.8 & 12.5\\
\cline{2-5}
 & \( 4 \to 3 \) & 44.4 & 24.3 & 6.1\\
\hline
\end{tabular}
\end{subtable}

\vspace{6pt}
\begin{subtable}{0.5\textwidth}
\centering
\caption{digits 5 and 7, \(B_{1}\)=39.2, \(B_{\infty}\)=0.3\label{table_1_j}}
\vspace{-6pt}
\setlength\extrarowheight{2pt}
\begin{tabular}{ |c|c|c|c|c| }
\hline
\multicolumn{2}{|c|}{\(\gamma\)} & 0.01 & 0.05 & 0.1\\
\hline
\multicolumn{2}{|c|}{\(C\)} & \(10^{4} \) & \(10^{4} \) & \(10^{4} \)\\
\hline
\multicolumn{2}{|c|}{Accuracy (\%)} & 99.3 & 99.6 & 99.1\\
\hline
\multirow{2}{*}{Succ rate (\%)} & \( 5 \to 7 \) & 48.8 & 24.2 & 25.7\\
\cline{2-5}
 & \( 7 \to 5 \) & 51.6 & 22.8 & 7.1\\
\hline
\end{tabular}
\end{subtable}
\end{table}

\begin{table}[h!]\small
\centering
\caption{Success rate of gradient-descent method evasion attacks
(binary class, USPS, with \(L_{1}\) and \(L_{\infty}\) limit)}
\label{table_1_usps_limit}

\vspace{-4pt}
\begin{subtable}{0.5\textwidth}
\centering
\caption{digits 0 and 1, \(B_{1}\)=76.8, \(B_{\infty}\)=0.6\label{table_1_k}}
\vspace{-6pt}
\setlength\extrarowheight{2pt}
\begin{tabular}{ |c|c|c|c|c|c| }
\hline
\multicolumn{2}{|c|}{\(\gamma\)} & 0.001 & 0.01 & 0.03 & 0.06\\
\hline
\multicolumn{2}{|c|}{\(C\)} & \(10^{4} \) & \(5 \times 10^{3} \) & \(5 \times 10^{3} \) & \(5 \times 10^{3} \)\\
\hline
\multicolumn{2}{|c|}{Accuracy (\%)} & 99.2 & 99.5 & 99.5 & 99.4\\
\hline
\multirow{2}{*}{Succ rate (\%)} & \( 0 \to 1 \) & 22.2 & 10.9 & 4.7 & 3.1\\
\cline{2-6}
 & \( 1 \to 0 \) & 100 & 100 & 29.5 & 8.8\\
\hline
\end{tabular}
\end{subtable}

\vspace{6pt}
\begin{subtable}{0.5\textwidth}
\centering
\caption{digits 2 and 3, \(B_{1}\)=51.2, \(B_{\infty}\)=0.3\label{table_1_l}}
\vspace{-6pt}
\setlength\extrarowheight{2pt}
\begin{tabular}{ |c|c|c|c|c|c| }
\hline
\multicolumn{2}{|c|}{\(\gamma\)} & 0.01 & 0.03 & 0.06 & 0.1\\
\hline
\multicolumn{2}{|c|}{\(C\)} & \(10^{5} \) & \(10^{5} \) & \(10^{5} \) & \(10^{5} \)\\
\hline
\multicolumn{2}{|c|}{Accuracy (\%)} & 97.8 & 97.8 & 97.8 & 97.8\\
\hline
\multirow{2}{*}{Succ rate (\%)} & \( 2 \to 3 \) & 50.3 & 34.2 & 25.4 & 20.2\\
\cline{2-6}
 & \( 3 \to 2 \) & 52.5 & 38.7 & 25.2 & 12.9\\
\hline
\end{tabular}
\end{subtable}

\vspace{6pt}
\begin{subtable}{0.5\textwidth}
\centering
\caption{digits 6 and 7, \(B_{1}\)=76.8, \(B_{\infty}\)=0.6\label{table_1_m}}
\vspace{-6pt}
\setlength\extrarowheight{2pt}
\begin{tabular}{ |c|c|c|c|c|c| }
\hline
\multicolumn{2}{|c|}{\(\gamma\)} & 0.01 & 0.02 & 0.03 & 0.06\\
\hline
\multicolumn{2}{|c|}{\(C\)} & \(10^{4} \) & \(10^{4} \) & \(10^{4} \) & \(10^{4} \)\\
\hline
\multicolumn{2}{|c|}{Accuracy (\%)} & 99.7 & 99.7 & 99.7 & 99.7\\
\hline
\multirow{2}{*}{Succ rate (\%)} & \( 6 \to 7 \) & 58.9 & 38.2 & 19.4 & 7.6\\
\cline{2-6}
 & \( 7 \to 6 \) & 80.8 & 52.1 & 26.0 & 8.2\\
\hline
\end{tabular}
\end{subtable}
\end{table}

\myparagraph{Inter-Class-SV Distance.}
Since \(\gamma\) controls how quickly the RBF kernel vanishes, we are motivated
to compare minimum (Euclidean) distance between each support vector of one class
\( (sv_{1i}) \) and all support vectors of the opposite class \( (sv_{2j}) \),
\ie \( MinDist(sv_{1i}) = \arg\min_{j}\,Distance(sv_{1i}, sv_{2j}) \). Our
intuition is that a larger \( \gamma \) suggests (1) a quicker drop of values
for both the kernel function and the gradient; (2) a wider gap between the two
classes. Both observations contribute to the lower success rate of the evasion
attack.

Figures~\ref{figure_4_a} and~\ref{figure_4_b} present the minimum distance between
support vectors of classes ``3'' and ``4'' in the MNIST dataset (due to similarities, 
we omit the other results). Observe that when \(\gamma\) first decreases from
0.5, the support vectors of the opposite class move further away---here
the corresponding model is still less vulnerable to evasion attack.
As \(\gamma\) continues to decrease the trend reverses, \ie the support
vectors of opposite class move closer to each other. A smaller \(\gamma\)
already means the RBF kernel vanishes more slowly, and the closer distance
between the two classes makes it even easier for a test sample to cross the
decision boundary. Consequently, the corresponding model becomes much more
vulnerable.

This prompts the question: Given a model with a \(\gamma\),
is there a way to determine whether the model is robust? The results in
Tables~\ref{table_1_a}--\ref{table_1_c},~\ref{table_1_e}--\ref{table_1_g} 
and~\ref{table_1_spambase} witness a strong correlation between
success rate and the percentage of ``\( 2/\sqrt{\gamma} \geq \) minimum
distance''---the lower the percentage the less vulnerable the model.

\myparagraph{Margin Explanation.}
We have also observed a positive correlation between margin per support vector
and the minimum distance calculated above. These findings suggest
that \textbf{separated inter-class support vectors lead to more secure models},
lending experimental support to the geometric argument (Section~\ref{sec:example}).

\myparagraph{Linear SVM.}
The experiments are performed for linear SVMs on MNIST, with results serving
as baseline. Table~\ref{table_1_d} demonstrates that success rates under
linear models are 100\%, as expected. However, a larger \(C\) requires smaller
changes to the target sample, as larger \(C\) leads to smaller margin.

\myparagraph{Discussion on model robustness vs. overfitting.}
It should be noted that we are not suggesting larger \(\gamma\) induces robustness. 
Instead, we demonstrate a phrase transition with small increase to \(\gamma\) causes 
significant success rate drop for evasion attacks. Further increase to \(\gamma\) 
over the threshold value offers little benefit. Our current results suggest that  
the model may (Table~\ref{table_1_spambase}) or may not (Tables~\ref{table_1_mnist}, 
~\ref{table_1_usps}) overfit when \(\gamma\) reaches the threshold value.

\subsection{Impact of \(\gamma\) on Vulnerability (Multiclass)}\label{sec:exp-grad-gamma}
This section further investigates the impact of \(\gamma\) on success rate of
evasion attacks, in multiclass scenarios. For the MNIST dataset, 
two RBF SVMs with \(\gamma\) as 0.05 and 0.5, are trained on both \(D_{1}\)
and \(D_{2}\), respectively. For comparison, four linear SVMs are also trained
on \(D_{1}\) and \(D_{2}\) with different values of \(C\). For the USPS dataset, 
three RBF SVMs are trained, where \(\gamma\) = 0.02, 0.1 and 0.5. An attack 
is still considered successful if the perturbed test sample is misclassified within 30 steps.

As can be seen from Tables~\ref{table_2_a} and~\ref{table_2_b}: (1) for RBF SVMs, the success rates
under the models with larger \(\gamma\) are much lower; (2) for linear SVMs
the success rates are always 100\%, but the average \(L_{1}\) change is
smaller as \(C\) increases---observations consistent with previous results.

\begin{table}[h!]\small
\centering
\caption{Success rate and average \(L_{1}\) change of gradient-descent method
evasion attack, in multiclass scenarios, MNIST\textsuperscript{a}}
\label{table_2_a}
\vspace{-6pt}
\begin{threeparttable}
\setlength\extrarowheight{2pt}
\begin{tabular}{ |c|c|c|c|c| }
\hline
\multicolumn{2}{|c|}{} & \parbox{1.2cm}{\centering Accuracy\\(\%)} & \parbox{1.2cm}{\centering Succ rate\\(\%)} & \pbox{1.6cm}{Ave \(L_{1}\)\\change}\\
\hline
\multirow{2}{2.3cm}{\centering RBF\, \( (\gamma=0.05, \, C=10^{3}) \)} & Model 1 & 87.2 & 91.8 & \(39.4^{b}\)\\
\cline{2-5}
& Model 2 & 87.7 & 92.8 & \(40.5^{b}\)\\
\hline
\multirow{2}{2.3cm}{\centering RBF\, \( (\gamma=0.5, \, C=10) \)} & Model 1 & 94.8 & 24.4 & \(17.7^{b}\)\\
\cline{2-5}
& Model 2 & 94.8 & 23.7 & \(18.2^{b}\)\\
\hline
\multirow{2}{2.3cm}{\centering Linear\\ \( (C=10^{3}) \)} & Model 1 & 89.0 & 100 & 28.5\\
\cline{2-5}
& Model 2 & 89.2 & 100 & 26.8\\
\hline
\multirow{2}{2.3cm}{\centering Linear\\ \( (C=2 \times 10^{4}) \)} & Model 1 & 91.4 & 100 & 18.7\\
\cline{2-5}
& Model 2 & 91.7 & 100 & 18.1\\
\hline
\end{tabular}
\begin{tablenotes}
\footnotesize
\item[a] Since it takes more than an order of magnitude longer to run experiments using RBF SVM than linear kernel, each result regarding RBF SVM is based on 1000 test samples, while each linear SVM result is based on 5000 test samples.
\item[b] Only the successful cases are counted.
\end{tablenotes}
\end{threeparttable}
\end{table}

\vspace{-6pt}
\begin{table}[h!]\small
\centering
\caption{Success rate and average \(L_{1}\) change of gradient-descent method
evasion attack, in multiclass scenarios, USPS\textsuperscript{a}}
\label{table_2_b}
\vspace{-6pt}
\begin{threeparttable}
\setlength\extrarowheight{2pt}
\begin{tabular}{ |c|c|c|c|c| }
\hline
\(\gamma\) & \(C\) & Accuracy (\%) & Succ rate (\%) & Ave \(L_{1}\) change\\
\hline
0.02 & \(10^{4}\) & 89.9 & 89.6 & \(42.5^{b}\)\\
\hline
0.1 & \(2 \times 10^{3}\) & 92.1 & 38.7 & \(40.7^{b}\)\\
\hline
0.5 & \(10^{2}\) & 95.0 & 38.2 & \(40.9^{b}\)\\
\hline
\end{tabular}
\begin{tablenotes}
\footnotesize
\item[a] Each result is based on 1500 test samples.
\item[b] Only the successful cases are counted.
\end{tablenotes}
\end{threeparttable}
\end{table}

\section{Experiments: Gradient-Quotient Method}\label{sec:exp-quot}
In this section we present experimental results establishing the effectiveness
of our proposed method for generating adversarial samples.
\subsection{Attacking SVMs \& RBF Networks Directly}
Recall that based on our new approach, a test sample \(\v{x}\) is updated as
\(\v{x}_{t+1}= \v{x}_t - \varepsilon_t \cdot \nabla (f_{1}(\v{x})/f_{2}(\v{x})) \),
where \(f_{1}(\v{x})\) and \(f_{2}(\v{x})\) are the scores for the top two
scoring classes for \(\v{x}\). In order to test whether this method is more
effective than the popular gradient-descent method, we run similar experiments
to Section~\ref{sec:exp-grad-gamma}, (1) for the MNIST dataset, one RBF SVM
\( (\gamma = 0.5, C = 10) \) and one linear SVM \( (C = 1000) \) are trained on
\(D_1, D_2,\ldots, D_5 \), respectively; (2) for the USPS dataset, the RBF SVM with 
\( \gamma = 0.5, C = 10^{2} \) is reused.

Comparing the results in Table~\ref{table_2_a} and Table~\ref{table_3}, 
we observe that the gradient quotient method is very effective against 
SVMs trained on the MNIST dataset: (1) for the RBF SVMs with
\(\gamma = 0.5 \), the success rates increase from around 24\% to a resounding
100\%. Moreover we have tested a wide range of values for \(\gamma\) from
0.01 through 10, with \textbf{resulting success rates always 100\% under our new
approach}; (2) the required \(L_1\) perturbation also decreases.

However, the new method achieves a success rate of only 71.2\% 
(still higher than the gradient descent method's 38.2\%)
against the RBF SVM trained on the USPS dataset. An examination of the 
log files reveals that in this case, the second highest score \(f_{2}\) is always 
negative, and hence decreasing the quotient of \(f_{1}/f_{2}\) may not work. 
Interestingly, such finding coincides with our previous conclusion that 
models with wider inter-class gap are more robust.

\myparagraph{Attacking RBF networks.}
Does the gradient quotient method work with other types of model? We further test it 
against a RBF network trained on \(D_{1}\), with 600 RBF neurons and a learning rate 
of 0.05. The results show that the success rate is 74.1\% (\(f_{2}\) is also negative 
in this case), which is slightly lower than the gradient descent method's 79.9\%.

In summary, the proposed gradient quotient method works most effectively when the  
scores for both the original and target classes are positive. In other situations, it 
performs at least similarly to the gradient descent method---none of our experimental 
results on all datasets and two types of models has shown any obvious inferiority.

\begin{table}[h]
\centering
\caption{Success rate and average \(L_{1}\) change of the evasion attack (the gradient quotient method, multiclass scenarios, MNIST).\textsuperscript{a}}
\label{table_3}\small
%\begin{adjustbox}{max width=.5\textwidth}
\vspace{-6pt}
\begin{threeparttable}
\setlength\extrarowheight{2pt}
\begin{tabular}{ |c|c|c|c|c| }
\hline
%\multicolumn{2}{|c|}{} & Accuracy(\%) & Suc rate(\%) & Ave\(L_{1}\)change\\
\multicolumn{2}{|c|}{} & \parbox{1.2cm}{\centering Accuracy\\(\%)} & \parbox{1.2cm}{\centering Succ rate\\(\%)} & \pbox{1.6cm}{Ave\(L_{1}\)\\change}\\
\hline
\multirow{5}{2cm}{\centering RBF\\ \( (\gamma=0.5,\) \\ \(C=10) \)} & Model 1 & 94.8 & 100 & 17.8\\
\cline{2-5}
& Model 2 & 94.8 & 100 & 17.9\\
\cline{2-5}
& Model 3 & 95.0 & 100 & 17.4\\
\cline{2-5}
& Model 4 & 95.2 & 100 & 17.5\\
\cline{2-5}
& Model 5 & 95.0 & 100 & 18.1\\
\hline
\multirow{5}{2cm}{\centering Linear\\ \( (C=10^3) \)} & Model 1 & 89.0 & 100 & 19.3\\
\cline{2-5}
& Model 2 & 89.2 & 100 & 20.6\\
\cline{2-5}
& Model 3 & 89.1 & 100 & 20.9\\
\cline{2-5}
& Model 4 & 89.2 & 100 & 20.8\\
\cline{2-5}
& Model 5 & 88.9 & 100 & 19.9\\
\hline
\end{tabular}
\begin{tablenotes}
\footnotesize
\item[a] Each result regarding RBF SVM is based on 800 test samples, while each result on linear SVM is based on 5000 test samples.
\end{tablenotes}
\end{threeparttable}
%\end{adjustbox}
\end{table}

\subsection{Attacking via Surrogate}
Up until now, we have implicitly assumed that the attacker possesses complete
knowledge of the target classifier, which may be unrealistic in practice. Hence,
we next examine attacks carried out via a surrogate. For example, in order to
mislead a RBF SVM, the attacker first trains their own RBF SVM on a similar
dataset, builds the attack path of how a test sample should be modified, then
applies it to the target SVM.

Since there is no guarantee that
the surrogate and target classifiers misclassify the test sample simultaneously,
all test samples are modified 15 times by the surrogate in this experiment; an
attack is considered to be successful if the target classifier misclassifies
the adversarial sample within 15 steps too. In addition, we modify the method as:
\hspace*{-1cm}
\[  \v{x}_{t+1} =
  \begin{cases}
	  \v{x}_t - \varepsilon_t \cdot \nabla (f_{1}(\v{x}_t)/f_{2}(\v{x}_t)) & \text{ prior to surrogate}\\
  & \text{ misclassification}\\
	  \v{x}_t + \varepsilon_t \cdot \nabla (f_{1}(\v{x}_t)/f_{2}(\v{x}_t)) & \text{ otherwise}
  \end{cases}
\]

In other words, before the test sample is misclassified by the surrogate, it
travels ``downhill'', but after crossing the decision boundary it travels
``uphill''. Otherwise the test case continues oscillating back and forth around
the boundary.

\myparagraph{Intra-model transferability.}
We reuse the RBF SVMs trained on \(D_1, D_2,\ldots, D_5 \), each of which serving as both
surrogate \((S_i)\) and target \((T_i)\) classifiers. As can be seen from
Table~\ref{table_4_rbf}, the success rates are all over 65\%. Specifically, those
values inside the bracket are the success rates when the target classifier
misclassifies before the surrogate, while the values outside are the overall
success rates. For comparison, the same experiments have been performed for
linear SVMs, producing similar success rates (54\%---71\%), \textbf{notably higher
than previous findings (around 40\%) reported} by
\citet{Papernot2016_transferability}.

\begin{table}[h]\small
\centering
\caption{Success rate of evasion attacks via surrogate (RBF SVM, MNIST).\textsuperscript{a}}
\label{table_4_rbf}
\vspace{-6pt}
\begin{threeparttable}
\setlength\tabcolsep{4pt}
\setlength\extrarowheight{2pt}
\begin{tabular}{ |c|c|c|c|c|c| }
\hline
& \(T_1\) & \(T_2\) & \(T_3\) & \(T_4\) & \(T_5\)\\
\hline
\(S_1\) & 100 & 66.1 (9.2) & 69.8 (9.5)	& 68.3 (9.8) & 66.4 (7.1)\\
\hline
\(S_2\) & 68.0 (8.2) & 100 & 67.2 (8.4)	& 72.4 (10.4) & 67.9 (7.4)\\
\hline
\(S_3\) & 65.0 (8.4) & 66.1 (9.4) & 100	& 67.3 (11.1) & 65.9 (8.5)\\
\hline
\(S_4\) & 65.3 (10.0) & 67.2 (10.8) & 67.0 (10.6) & 100 & 65.3 (8.8)\\
\hline
\(S_5\) & 67.0 (9.2) & 67.8 (8.9) & 69.4 (9.0) & 69.4 (9.1) & 100\\
\hline
\end{tabular}
\begin{tablenotes}
\footnotesize
\item[a] Each result is based on 800 test samples.
\end{tablenotes}
\end{threeparttable}
\end{table}

%\begin{table}[h]\small
%\centering
%\caption{Success rate of evasion attacks via surrogate (Linear SVM, MNIST).\textsuperscript{a}}
%\label{table_4_linear}
%\vspace{-6pt}
%\begin{threeparttable}
%\setlength\extrarowheight{2pt}
%\begin{tabularx}{0.5\textwidth}{ @{\extracolsep{\fill}}|C|C|C|C|C|C| }
%\hline
%& \(T_1\) & \(T_2\) & \(T_3\) & \(T_4\) & \(T_5\)\\
%\hline
%\(S_1\) & 100 & 59.5 & 54.4 & 59.3 & 55.7\\
%\hline
%\(S_2\) & 65.4 & 100 & 60.4 & 63.9 & 59.7\\
%\hline
%\(S_3\) & 64.1 & 65.5 & 100 & 64.8 & 62.3\\
%\hline
%\(S_4\) & 67.3 & 66.1 & 60.8 & 100 & 60.4\\
%\hline
%\(S_5\) & 70.6 & 66.7 & 61.4 & 67.0 & 100\\
%\hline
%\end{tabularx}
%\begin{tablenotes}
%\footnotesize
%\item[a] Each result is based on 4000 test samples.
%\end{tablenotes}
%\end{threeparttable}
%\end{table}

\myparagraph{Inter-model transferability.}
Is gradient quotient based evasion attack still effective if the surrogate and target models 
are of different types? We keep the above five RBF SVMs as surrogate, but change the target 
classifier to the RBF network (trained on \(D_{1}\)) from the last subsection. The results 
show that the success rates range from 61.3\% to 69.1\%, which are very close to the values 
in Table~\ref{table_4_rbf}.

\subsection{Mimicry Attacks}
The previous two subsections have demonstrated that our new approach is effective in 
manipulating test samples. But what if the attacker intends to make the test sample 
misclassified as a specific class?

In order to test the gradient quotient method for mimicry attacks, we make the following change: 
\(\v{x}_{t+1}= \v{x}_t - \varepsilon_t \cdot \nabla (f_{1}(\v{x})/f_{T}(\v{x})) \), 
where \(f_{T}(\v{x})\) corresponds to the score of the target class, while \(f_{1}(\v{x})\) is 
still the score for the top scoring class.

\begin{figure}
\centering
\begin{subfigure}{0.5\textwidth}
  \centering
  \includegraphics[width=\textwidth]{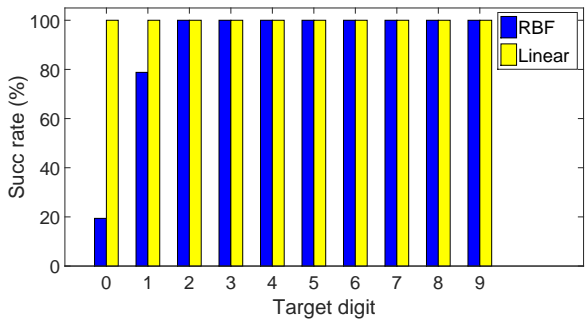}
  \caption{Success rate}
  \label{figure_mimicry_rate}
\end{subfigure}

\begin{subfigure}{0.5\textwidth}
  \centering
  \includegraphics[width=\textwidth]{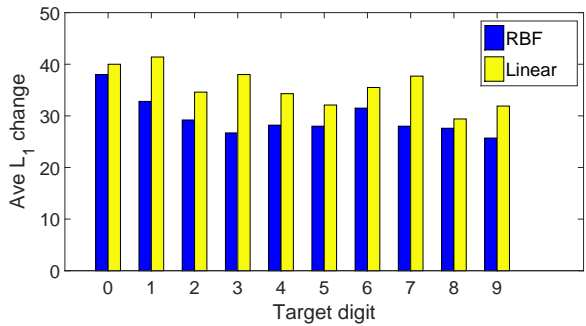}
  \caption{\(L_{1}\) change}
  \label{figure_mimicry_l1}
\end{subfigure}
\caption{Success rate and \(L_{1}\) change of mislabelling other digits as the target digit.}
\label{figure_mimicry}
\end{figure}

For each test sample in the MNIST dataset, we apply the gradient quotient algorithm to check 
whether it can be misclassified as each of the other nine digits. The RBF SVM \((\gamma = 0.5, C = 10)\) 
and the linear SVM (\(C\) = 1000) trained on \(D_{1}\) are reused here. The results in Figure~\ref{figure_mimicry} 
show that in most cases this approach can successfully make the original digit misclassified 
as the target. However, under the RBF SVM, the success rate is very low when the attack is to 
make other digits look like ``0''. The reason is still that \(f_{0}(\v{x})\), \ie the score for 
digit ``0'', is always negative.

Another interesting point is that among the successful cases where other digits are misclassified 
as ``0'', over half of them are digit ``6''. Therefore, we tried an indirect method---first modify 
the test sample so that it is considered as ``6'' (the success rate for this step is 100\%), and then 
follow the original method, i.e., modify ``6'' until either it is classified as ``0'', 
or the maximum step limit of 30 is reached. This indirect approach doubled the overall success rate.

%\begin{table}[h]\small
%\centering
%\caption{Success rate and \(L_{1}\) change of mislabelling other digits as the target digit.\textsuperscript{a}}
%\label{table_mimic}
%\vspace{-6pt}
%\begin{threeparttable}
%\setlength\extrarowheight{2pt}
%\begin{tabularx}{0.5\textwidth}{ @{\extracolsep{\fill}}|C|C|C|C|C| }
%\hline
%\multirow{2}{*}{Target digit} & \multicolumn{2}{c|}{Succ rate (\%)} & \multicolumn{2}{c|}{Ave \(L_{1}\) change}\\
%\cline{2-5}
%& RBF & Linear & RBF & Linear\\
%\hline
%0 & 19.4 & 100 & 38.0\textsuperscript{b} & 40.0\\
%\hline
%1 & 78.8 & 100 & 32.8\textsuperscript{b} & 41.4\\
%\hline
%2 & 100 & 100 & 29.2 & 34.6\\
%\hline
%3 & 100 & 100 & 26.7 & 38.0\\
%\hline
%4 & 100 & 100 & 28.2 & 34.3\\
%\hline
%5 & 100 & 100 & 28.0 & 32.1\\
%\hline
%6 & 100 & 100 & 31.5 & 35.5\\
%\hline
%7 & 100 & 100 & 28.0 & 37.7\\
%\hline
%8 & 100 & 100 & 27.6 & 29.4\\
%\hline
%9 & 100 & 100 & 25.7 & 31.9\\
%\hline
%\end{tabularx}
%\begin{tablenotes}
%\footnotesize
%\item[a] Each result regarding RBF SVM is based on 1500 test samples, while each result on linear SVM is based on 4000 test samples
%\item[b] Only the successful cases are counted
%\end{tablenotes}
%\end{threeparttable}
%\end{table}

\section{Conclusions and Future Work}\label{sec:conc}
Recent studies have shown that it is relatively easy to fool machine learning
models via adversarial samples. In this paper, we demonstrate that the
gradient-descent method---the leading approach to generate adversarial
samples---has limitations against RBF SVMs, when the precision parameter
\(\gamma\) controlling kernel smoothness is chosen properly. We find 
predictable phase transitions of attack success occur at thresholds that
are functions of geometric margin-like quantities measuring inter-class
support vector distances. Our characterisation can be used to make RBF
SVM more robust against common evasion and mimicry attacks.

We propose a new method for manipulating target samples into 
adversarial instances, with experimental results showing that this new
method not only increases attack success rate, but decreases the required
changes made to input points.

For future work, (1) regarding the gradient-descent method, we intend to
replicate and expand findings for \(\gamma\) and smoothness in general, in
other settings and for other classifiers. (2) We will continue exploring suitability
of our new generation approach when the target is not an SVM, with direct
attacks or SVM surrogates. (3) Further investigation into light-weight
yet efficient countermeasures also serves as an important direction for
future work.

\bibliographystyle{ACM-Reference-Format}
\bibliography{mylib} 

\end{document}